\documentclass{ws-ijmpd}
\usepackage{subfigure}
\usepackage{cite}
\usepackage{hyperref}
\usepackage{multirow}

\newcommand{\ba}{\begin{eqnarray}}
\newcommand{\ea}{\end{eqnarray}}
\newcommand{\be}{\begin{equation}}
\newcommand{\ee}{\end{equation}}
\newcommand{\bd}{\begin{displaymath}}
\newcommand{\ed}{\end{displaymath}}
\newcommand{\een}{\nonumber\end{equation}}
\newcommand{\bea}{\begin{eqnarray}}
\newcommand{\eean}{\nonumber\end{eqnarray}}
\newcommand{\eea}{\end{eqnarray}}

\begin{document}

\markboth{Karl Jansen}
{Cosmological Phase transitions from Lattice Field Theory}

\catchline{}{}{}{}{}
%
%%%%%%%%%%%%%%%%%%%%%%%%%%%%%%%%%%%%%%%%%%%%%%%%%%%%%%%%%%%%%%%%%%%%

\title{
\vspace*{-1.5cm}
\begin{flushright}
DESY 11-222
\end{flushright}
\vspace*{1.5cm}
COSMOLOGICAL PHASE TRANSITIONS FROM LATTICE FIELD THEORY}

\author{Karl Jansen}

\address{NIC, DESY, Platanenallee 6, 15738 Zeuthen, Germany}

\maketitle

%\begin{history}
%\received{Day Month Year}
%\revised{Day Month Year}
%\comby{Managing Editor}
%\end{history}

\begin{abstract}
In this proceedings contribution we discuss the fate of the 
electroweak and the quantum chromodynamics phase transitions 
relevant for the early stage of the universe
at non-zero temperature. These phase transitions are related 
to the Higgs mechanism and the breaking of chiral symmetry, 
respectively. We will review that non-perturbative lattice
field theory simulations show that these phase transitions
actually do not occur in nature and that physical observables 
show a completely smooth behaviour as a function of the 
temperature.  
\end{abstract}

\keywords{Electroweak Phase transition, QCD phase transition, Lattice Field Theory}

\section{Introduction}

During its early evolution, the universe is believed to have passed 
through two phase transitions. The first is the 
electroweak phase transition at temperatures of the electroweak scale of 
about 250GeV. The second is the quantum chromodynamics (QCD) 
phase transition at much later 
times and correspondingly smaller temperatures 
of $O(100)$MeV. 

Both phase transitions are associated with important phenomena related
to spontaneous symmetry breaking. In case of the electroweak theory, 
it is the Higgs mechanism providing masses to all quarks, 
leptons 
and the weak gauge bosons. In case of QCD, it is the breaking of chiral 
symmetry which is an essential element in understanding of  
hadronic phenomena in nature such as the hadron spectrum. 

Phase transitions are non-perturbative phenomena and hence need appropriate
tools to be investigated theoretically. Lattice field theory is such a tool 
and indeed, in the past both, the electroweak and the QCD non-zero temperature phase
transitions have been studied by means of numerical simulations. The picture
that emerged from these simulations is rather surprising: it seems that 
nature has preferred to evolve the universe {\em completely smoothly
without ever passing the anticipated phase transitions described
above.} 
Thus, the universe has merely changed its state of aggregation
like water does in the temperature-pressure phase diagram 
from the fluid (water) to the gaseous (vapor) regions of the phase diagram. 
In this 
proceedings contribution we will review the evidence that has been 
obtained for this picture 
from lattice simulations.
Although the results summarized here are well known in large 
parts of the high energy physics community, they might be 
new and interesting to the participants of this 
conference which provides sufficient motivation to discuss them here. 

\section{The non-zero temperature electroweak phase transition}

In the electroweak sector of the standard model our present 
understanding is that through the Higgs mechanism mass has been given 
to the electroweak gauge bosons, the quarks and the leptons. 
In the limit of pure scalar $\Phi^4$ theory, 
the Higgs mechanism is based
on the spontaneous breaking of an $O(4)$ symmetry of the action
that left behind the Higgs boson and three Goldstone bosons 
which then turned into massive gauge bosons through the interaction
of the Higgs field 
when the SU(2) gauge fields are switched on. 

The phenomenon of spontaneous symmetry breaking is associated 
with a phase transition and hence the universe should have undergone 
a non-zero temperature phase transition at which the Higgs mechanism 
became operative. 
Although the investigation of this scenario is clearly interesting
in its own right, large scale simulations of the non-zero temperature electroweak 
phase transition were started in the context of the question, whether
in the standard model of particle interactions the Sakharov 
conditions for explaining the baryon asymmetry of the universe
can be realized. One of Sakharov's condition is that the universe
had to be sufficiently out of thermal equilibrium in order that 
an asymmetry between baryon anti-baryon generation and annihilation 
processes can occur. Sufficient here means that the ratio 
of the scalar field 
expectation value $v$ over the critical temperature $T_c$ is larger than one, 
$v/T_c > 1$. This argument can be inferred from the sphaleron 
transition rate and we refer to ref.~\cite{Cohen:1993nk}\footnote{In this short
proceedings contribution we cannot provide a comprehensive reference list. 
We point therefore to only selected references and reviews.} for more
details. 

Although there were a number of first attempts within perturbation theory 
to clarify this question, it soon turned out that these early attempts 
were not very appropriate to describe the phase transition. 
This triggered in turn numerical simulations of the electroweak 
sector of the standard model both in the four dimensional system 
\cite{Bunk:1992kf,Fodor:1994dm,Fodor:1994sj,Csikor:1996sp}
and 3-dimensional effective theories
\cite{Kajantie:1995kf,Kajantie:1996mn}. 
In all these simulations
the fermions were neglected (the value of the Yukawa coupling
were set to zero) such that pure SU(2)-Gauge-Higgs systems were
considered. 

\begin{figure}
%\hspace{0.5cm}
\includegraphics[width=0.45\textwidth]{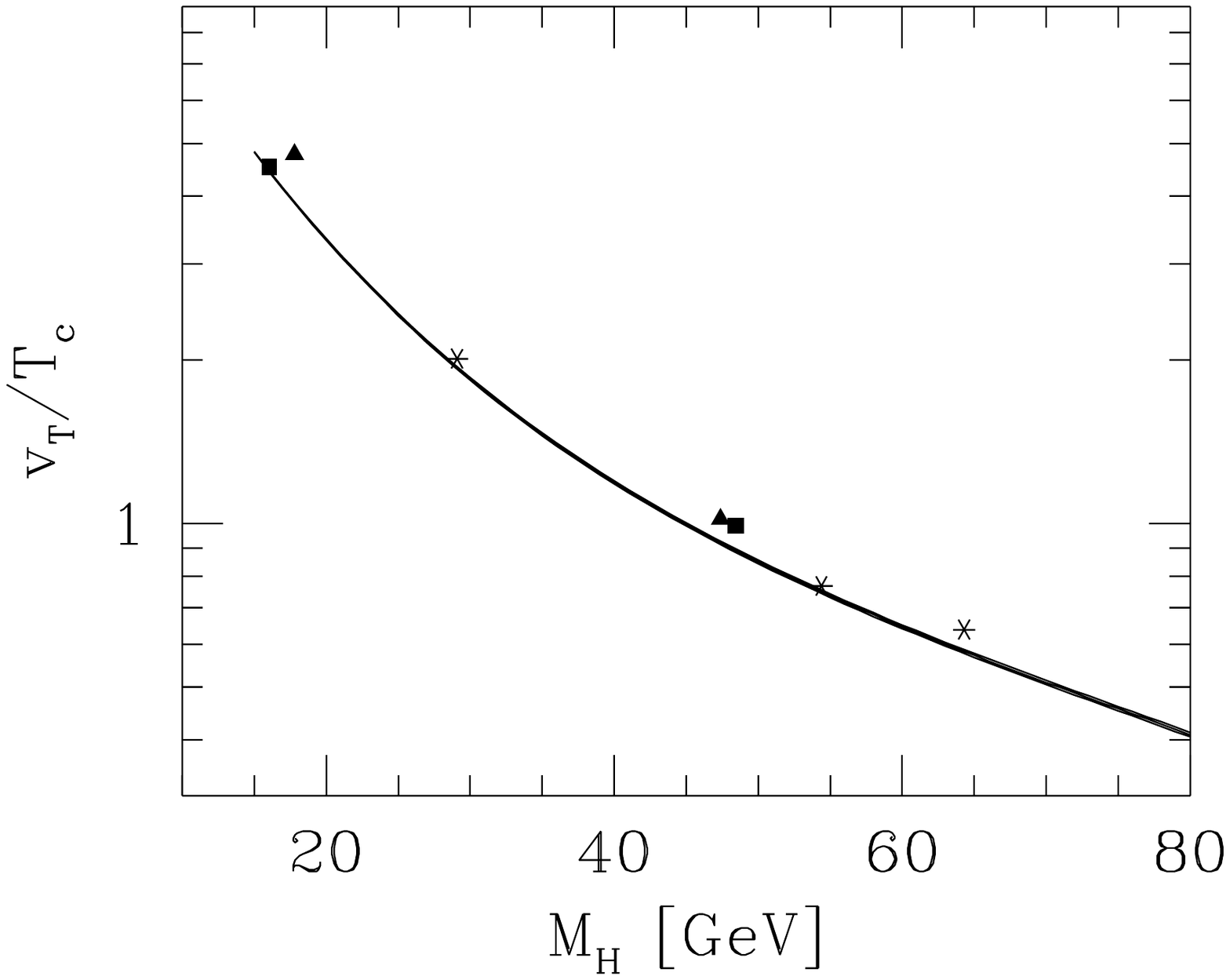}
%\vspace*{-3cm}
\hspace*{1.0cm}\includegraphics[width=0.42\textwidth]{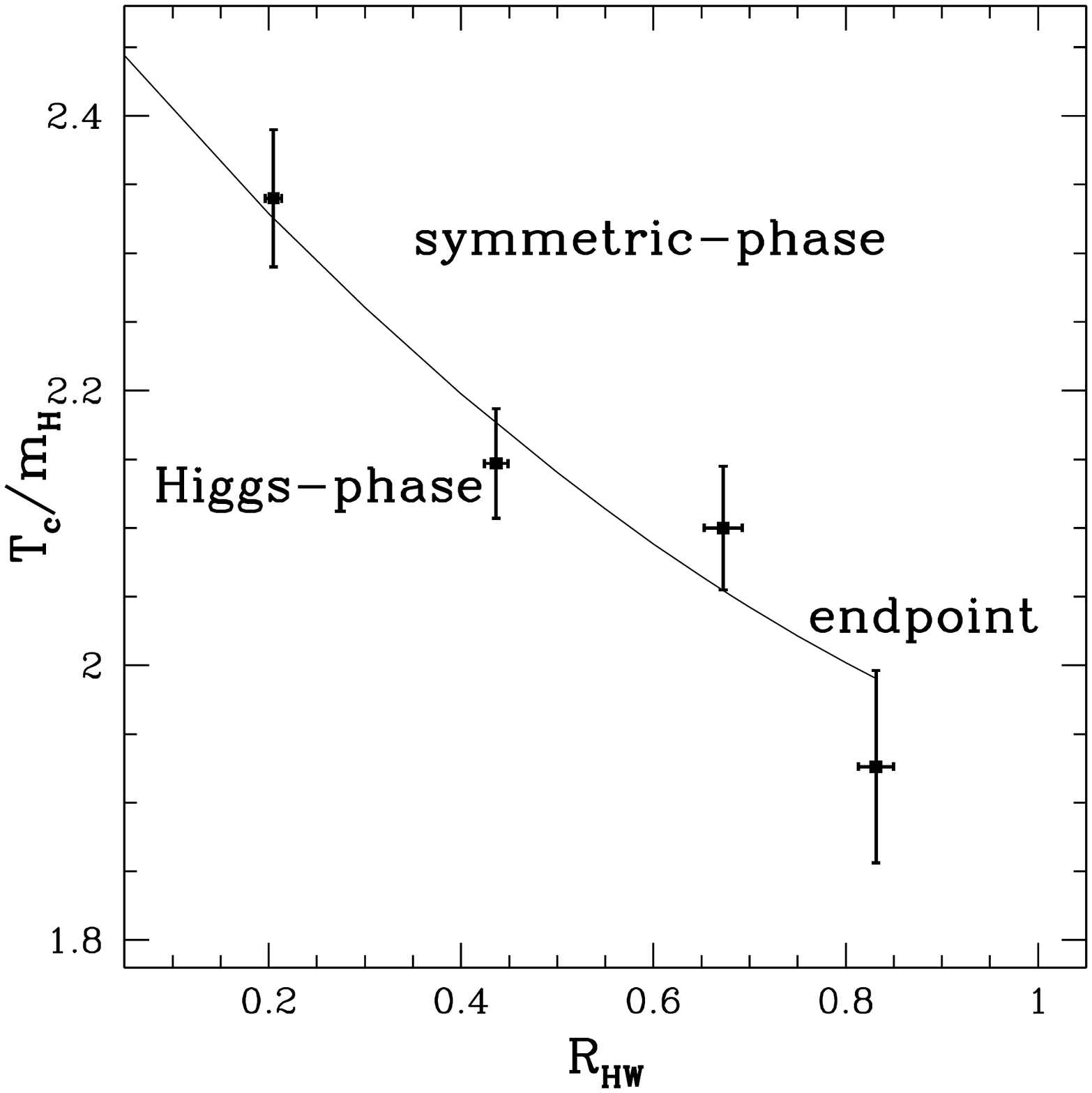}
\caption{Left: the ratio of the electroweak vacuum expectation value 
$v$ over the critical temperature $T_c$ is shown as a function of the 
Higgs boson mass. The data points are from simulations of
the 4-dimensional theory and effective 3-dimensional models. 
The solid line is from a 2-loop perturbative analysis 
of the gauge invariant effective potential.  
The figure is taken from ref.~\cite{Jansen:1995yg}.
Right: we show the ratio of the critical temperature over the Higgs boson mass 
as a function of the ratio $R_{\rm HW}$ of the 
Higgs boson over the W-boson mass. 
At a value of $R_{\rm HW} \approx 0.8$ the phase transition ends and 
turns into a crossover, demonstrating that for Higgs boson masses larger 
than about 100GeV no phase transition in the electroweak sector 
of the standard model has occurred. 
\label{vttc}}
\end{figure}

From these simulations it then became possible to compute 
$v/T_c$ as a function of the Higgs boson mass $M_{\rm H}$. The results 
of such simulations and a comparison to a refined perturbative
approach \cite{Buchmuller:1994vy,Buchmuller:1995sf} 
are summarized in ref.~\cite{Jansen:1995yg} and 
shown in fig.~\ref{vttc} (left).   
Note that in these works which were performed in a completely
gauge invariant setup the vacuum expectation value was defined as
$v^2={\rm Tr} \langle \Phi^\dagger_x\Phi_x\rangle$ with 
$\Phi_x$ the Higgs field. 
It was found that for Higgs boson masses 
$M_{\rm H} \lesssim 80$GeV indeed a phase transition 
of first order can be established. However, for larger values
of $M_{\rm H}$ this phase transitions ends in a critical point
and turns into a cross-over, see fig.~\ref{vttc} (right). 
This means that physical observables 
depend completely smoothly on the temperature and do not feel
any phase transition. 
Since the present lower limit on the Higgs boson mass is  
$M_{\rm H} > 114$GeV, as a consequence 
--at least within the framework of the standard model--
the universe 
must have moved in temperature in the cross-over region and 
hence has not felt the presence of a phase transition. 
This conclusion from lattice simulations is supported by 
perturbative calculations of the effective potential 
{\em in a gauge invariant manner} up to two loops 
\cite{Buchmuller:1994vy,Buchmuller:1995sf} and 
from a non-perturbative renormalization group 
approach \cite{Bergerhoff:1994sj}. 

\section{The non-zero temperature phase transition of \\ quantum chromodynamics}

Let us now turn to the case of the strong interaction, described
theoretically within quantum chromodynamics in the framework 
of the standard model. Here it is expected that another kind 
of spontaneous symmetry breaking has occurred, namely 
chiral symmetry breaking. This phenomenon in QCD is considered
to be a most important aspect of the theory since it leads 
to the formation of a quark condensate and explains the 
smallness of the masses of the pions when compared to other 
hadron masses. The reason for this smallness is simply the 
association of the pions with the Goldstone bosons that 
emerged from the spontaneous breaking of chiral symmetry. 

\begin{figure}
%\vspace{-1cm}
\hspace{0.5cm}
\includegraphics[width=1.0\textwidth]{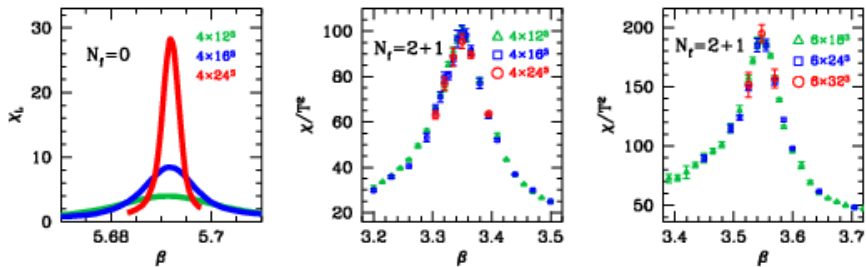}
\vspace{1cm}
\caption{We show the Polyakov loop susceptibility as a function 
of the temperature (encoded in the gauge coupling $\beta$) 
at fixed value of the temporal and various spatial extents of 
the lattice.
Case (a) corresponds to the situation when quarks are not 
considered as dynamical degree of freedom, leftmost graph, (b)
for the case of dynamical quarks with fixed time extent  
of the lattice $L_t=4$, middle and $L_t=6$ rightmost graph, taken 
from ref.~\cite{Fodor:2009ax}. 
\label{qcdpt}}
\end{figure}

Again, it is expected that chiral symmetry is associated 
with a phase transition that occurred in the early phase of 
the universe at a temperature of $O(100)$MeV. The determination of 
the critical temperature of the phase transition and its nature 
has been a major research direction in lattice QCD. 
The phase transition was first studied neglecting the 
quarks as dynamical degrees of freedom and indeed, a clear
signal of the non-zero temperature phase transition could be 
established, see fig.~\ref{qcdpt} left most graph. 
Here, the Polyakov loop susceptibility is plotted 
as a function of the inverse bare gauge coupling $\beta$ corresponding
to changing the temperature.
The susceptibility shows a peak behaviour with 
an increase of the peak height and a narrowing of the 
peak structure when the spatial lattice size is increased. 
This behaviour strongly indicates that in the infinite volume 
limit there is indeed a phase transition. 

When the quarks are turned on as dynamical degrees of freedom 
in the simulations with masses close to the physical 
value of the pion mass, the picture changes, however, completely. 
This can be seen in 
the middle and rightmost graphs in fig.~\ref{qcdpt}. For a fixed 
value of the time extent of the lattice,   
the susceptibilities 
do not grow at all with increasing spatial lattice size, 
but fall on top of each other for all 
lattice sizes used. This means that in the infinite volume limit 
there is no phase transition but just a cross-over. Since 
the simulations were performed at about the physical value of the 
pion mass \cite{Aoki:2006we,Aoki:2009sc,Borsanyi:2010bp}
this means that again all physical observables
behave completely smoothly as function of the temperature 
and do not feel any presence of a phase transition. 
For other results and reviews, see ref.~\cite{Cheng:2009be}
and refs.~\cite{Kanaya:2010qj,Fodor:2009ax}, respectively. 
Thus, nature has --as in case of the electroweak phase transition--
decided to not let the universe pass through a phase transition 
during its evolution.

\section{Discussion}

The results of the previous sections leave us with a puzzle. 
In the introduction
we have emphasized that the phenomenon of spontaneous symmetry 
breaking, associated with a phase transition, 
is at the heart of the standard model. It is the basis of 
the Higgs mechanism in the framework of the electroweak theory 
and chiral symmetry breaking in the case of quantum chromodynamics. 
The non-perturbatively obtained results from lattice field theory 
investigations tell us, however, that for a physical pion mass 
and experimentally allowed values of the Higgs boson mass the universe has not undergone 
any phase transition when cooling down. 

%\noindent {\em Electroweak sector}

In the electroweak sector of the standard model it is actually
not so surprising that there is no phase transition.
It is known since a long time that the symmetric regime and the 
broken regime of the electroweak 
theory are analytically connected 
\cite{Fradkin:1978dv} and in fact that they are one and the same phase.
Indeed, for the case of zero values of the Yukawa couplings there is even 
an analytical proof of this statement \cite{Seiler:1982pw}. 

Nevertheless, a ``Higgs mechanism'' is operative in the theory in the sense
that one finds massive gauge bosons and Higgs boson masses. 
In the gauge invariant setting of lattice computations, 
the Higgs field $\Phi_x$ is unphysical and has to be replaced
by local, composite and gauge invariant operators.
These operators can then be used to construct suitable 
correlation functions from which, through a transfer matrix 
decomposition, the desired masses are extracted. 
Examples for such composite operators are  
${\rm Tr}[\Phi^\dagger_x\Phi_x]$ for the Higgs boson 
and ${\rm Tr}[\Phi^\dagger_x U_{x,\mu}\Phi_{x+\mu}\tau]$, $\tau$ being a Pauli matrix,
for the case of the vector bosons.
The analysis of these correlation functions provides then 
the spectrum of the theory \cite{Evertz:1985fc,Montvay:1984wy} which 
is completely consistent with the expectation from 
the Higgs mechanism.

%\noindent {\em Quantum chromodynamics}

Also in the case of QCD, nature seems to have preferred a continuous behaviour
of physical observables as a function of the temperature. At very high 
temperatures  
the quarks 
and the gluons were immersed in a 
plasma characteristic of the high temperature regime 
of QCD. 
When the universe evolved, at a temperature of about $O(100)$MeV  
the coupling grew strong, the scalar 
condensate increased significantly in value and the formation of bound states occurred. 
All this happened in a completely smooth fashion without the need
of explicitly going through a phase transition, as already discussed    
in ref.~\cite{Wilczek:2006gk}. 
``Chiral symmetry breaking'' is then operative in the sense
that the observed hadron spectrum is obtained with the 
pions having much lighter masses than all other hadrons. 

%\noindent {\em Remark}

The purpose of this proceedings contribution has been 
to emphasize that in both, the electroweak and 
the strong sector of the standard model 
no phase transition occurred in the early universe. 
This conclusion is obtained from lattice simulations
in a completely gauge invariant setup. 
The lessons from this observation may be the following. 
\begin{itemize}
\item First, the Higgs mechanism of the electroweak theory and 
the chiral transition in QCD can both be 
described in a fully gauge invariant manner. 
%\item It seems that there is no need for any explicit symmetry breaking.
\item By choosing the masses of the Higgs
boson 
and the quarks to be large enough, 
in the evolution of the universe 
the phase transitions were avoided maybe just in order to not 
destroy physical phenomena, such as the baryon asymmetry, 
that were generated at a very early stage of the 
universe. 
\item The fact that the baryon asymmetry of the universe cannot be 
explained within the standard model is, of course, another 
manifestation of the incompleteness of the standard model 
neccessating  its replacement by a new, so far unknown new physics. 
\end{itemize}
In the end, it is however unclear, whether these observations 
provide  
non-trivial insight into the structure of the standard model, 
or, whether they are just the gauge invariant description  
of the phenomenon 
of spontaneous symmetry breaking.

\section*{Acknowledgments}
The author thanks the organizers of SMFNS2011 in Varadero, Cuba 
for giving him
the opportunity to discuss the topic of cosmological phase transitions
from lattice simulations in a very friendly and supporting 
atmosphere. I am also indebted to G.~Rossi and M.~Testa for 
interesting discussions. 

\bibliographystyle{unsrt_nt}
\bibliography{paper}

\end{document}